\documentclass[12pt]{article}
\pdfoutput=1

\usepackage{amssymb,amsmath,bm}
\usepackage{cite}
\usepackage{graphicx}
\usepackage{color}

\def\slash#1{\setbox0=\hbox{$#1$}\dimen0=\wd0
      \setbox1=\hbox{/} \dimen1=\wd1 \ifdim\dimen0>\dimen1
      \rlap{\hbox to \dimen0{\hfil/\hfil}} #1                        \else
      \rlap{\hbox to \dimen1{\hfil$#1$\hfil}}
      /   \fi}

\newcommand{\lsim}{
\mathrel{\hbox{\rlap{\hbox{\lower4pt\hbox{$\sim$}}}\hbox{$<$}}}}

\newcommand{\gsim}{
\mathrel{\hbox{\rlap{\hbox{\lower4pt\hbox{$\sim$}}}\hbox{$>$}}}}

\allowdisplaybreaks[2]

\long\def\symbolfootnote[#1]#2{\begingroup%
\def\thefootnote{\fnsymbol{footnote}}\footnote[#1]{#2}\endgroup} 

\begin{document}
\begin{titlepage}
\vspace*{-1.0truecm}

\begin{flushright}
TUM-HEP-773/10\\
revised version
\end{flushright}

\vspace{0.6truecm}

\begin{center}
\boldmath
{\Large\textbf{See-Saw Masses\\[0.3em] for Quarks and Leptons
  in $SU(5)$}}
\unboldmath
\end{center}

\vspace{0.4truecm}

\begin{center}
{\bf Thorsten Feldmann}\symbolfootnote[1]{Address since January 2011: \newline
IPPP, Department of Physics, University of Durham, Durham DH1 3LE, UK. \newline
email:thorsten.feldmann@durham.ac.uk}
\vspace{0.4truecm}

{\footnotesize
{\sl Physik Department, Technische Universit\"at M\"unchen, \\
James-Franck-Stra{\ss}e, D-85748 Garching, Germany}\vspace{0.2truecm}
}

\end{center}

\vspace{0.5cm}
\begin{abstract}
\noindent

We build on a recent paper by Grinstein, Redi and Villadoro,
where a see-saw like mechanism for quark masses was derived
in the context of spontaneously broken gauged flavour
symmetries. The see-saw mechanism is induced by heavy
Dirac fermions which are added to the Standard Model
spectrum in order to render the flavour symmetries 
anomaly-free. In this letter we report on the embedding
of these fermions into multiplets of an $SU(5)$ 
grand unified theory and discuss a number of interesting
consequences.

\end{abstract}

\end{titlepage}
\setcounter{page}{1}
\pagenumbering{arabic}


\section{Introduction} \label{sec:intro}

The phenomenological motivations for considering extensions of the Standard Model (SM) which
are based on grand unified theories (GUTs; see, for instance, 
\cite{Pati:1973uk,Georgi:1974sy,Carlson:1975gu,Fritzsch:1974nn,Mohapatra:1974gc},
or 
\cite{Buras:1977yy,Slansky:1981yr,Senjanovic:2006nc,Raby:2006sk}
 and references therein) are two-fold:
\begin{itemize}
 \item First, the embedding of the SM gauge group $SU(3)\times SU(2)\times U(1)$
  into simple or semi-simple gauge groups like $SU(4) \times SU(2) \times SU(2)$, $SU(5)$, or $SO(10)$,
  to name a few popular examples, allows for a reduction of independent gauge coupling 
  constants. Indeed, from the renormalization-group (RG) running of the measured SM
  couplings, there is evidence that a unification around scales of $\sim 10^{15}~{\rm GeV}$
  is possible \cite{Amaldi:1991cn}, although for this idea to work quantitatively, additional
  degrees of freedom at high energies should contribute to the RG running. 
  A well-known example which could exhibit the unification of
  coupling constants is, of course, the minimally super-symmetric SM (MSSM). 

 \item Second, the embedding of the SM fermions 
   into \emph{a few} (smaller) GUT representations appears attractive, both, for 
   minimalistic reasons as well as for a possible reduction of independent mass and mixing
   parameters in the flavour sector. Except for the similar masses of the $\tau$-lepton and
   the $b$-quark, there is, however, no direct evidence for such simplifications.
   On the other hand, with a suitable Higgs sector -- sometimes also supplied with 
   additional constraints from postulated discrete flavour symmetries -- reasonable
   quark and lepton spectra can be obtained from GUT models. Often GUT models also
   allow for naturally tiny neutrino masses through a see-saw mechanism, e.g.\ 
   induced by heavy Majorana neutrinos.

\end{itemize}

In a slightly different context, Grinstein, Redi and Villadoro \cite{Grinstein:2010ve}
have recently proposed an approach to generate the SM  Yukawa couplings 
\emph{in the quark sector} from a see-saw
mechanism, which is induced by heavy Dirac fermions that appear in constructions with
gauged flavour symmetries at (relatively) low energies (for related work,
see also \cite{Albrecht:2010xh,Feldmann:2009jung} and references
therein). 

In this letter, we are going to show that the mechanism in \cite{Grinstein:2010ve}
can be modified and extended in such a way that up-type and down-type quarks, 
charged leptons and neutrinos, as well as their heavy partners can be embedded into 
multiplets of $SU(5)$. The fermion spectrum below the GUT scale can be represented 
in an anomaly-free way with respect to a flavour symmetry group of the SM gauge interactions. 
In contrast to the original approach in \cite{Grinstein:2010ve}, 
our modification also introduces new heavy $SU(2)_L$ doublets, and therefore, in principle, 
offers the potential to adjust the unification of SM coupling constants via the
RG running and threshold corrections induced by
the new fermions. Furthermore, it can lead to a natural suppression of
neutrino masses with respect to the masses of the charged leptons.
Finally, the $SU(5)$ fermion spectrum can be made anomaly-free with respect
to the flavour symmetry of the GUT theory by introducing a set of heavy fermions
that are distinguished from the SM matter by a postulated $Z_2$-symmetry and
contain a SM singlet component which may serve as a dark matter
candidate.

\section{See-Saw mechanism with Heavy Dirac Fermions}

The inclusion of heavy Dirac partners
for the SM particles is motivated by the requirement that the chiral anomalies
related to the flavour symmetries of the SM gauge sector are canceled between
SM quarks and their heavy partners, which would allow for a promotion of the global flavour symmetries
to local transformations of a renormalizable gauge theory. This, in particular, removes the Goldstone 
modes that appear when the flavour symmetries are to be spontaneously broken by the vacuum expectation
values (VEVs) of some scalar fields.
The choice of quantum numbers for such fermion fields is not completely fixed:
The model in \cite{Grinstein:2010ve} contains the minimal set of fermions 
needed to reproduce the SM quark Yukawa terms from renormalizable dim-4 operators, only.
The approach discussed in \cite{Albrecht:2010xh} already starts from an effective theory
where the heavy fermions decouple from the SM fermions. 
The model that we are going to discuss in this work
represents a modification of \cite{Grinstein:2010ve},  based
on dim-4 terms with different fermion content and couplings. 

Let us first summarize the relevant expressions in the
Lagrangian relevant for energies \emph{below} the GUT scale.

\subsection{Down-Quark Sector}

The down-quark sector of our model coincides with the 
construction in \cite{Grinstein:2010ve}. It features a dim-4
Lagrangian
\begin{align}
 {\cal L} \ \ni \ & \bar Q_L \, H \, \psi_{d_R} + \bar \psi_{d} \, S_D \, \psi_{d_R} + M_D \, \bar \psi_{d}  D_R + \mbox{h.c.}
\label{dim4d}
\end{align}
The notation for the new fermion fields $\psi_{d_R}$ and $\psi_d$ follows  \cite{Grinstein:2010ve}.
Here and in the following, we will not specify ${\cal O}(1)$ coupling constants for simplicity.
The explicit Dirac mass $M_{D}$ controls the coupling between the right-handed SM quarks
with their heavy partners. $S_{D}$ is a matrix-valued scalar field, which we will refer to as a spurion field
as in the context of minimal flavour violation \cite{D'Ambrosio:2002ex}. It will achieve a VEV from an (unspecified)
potential. Finally, the VEV of the SM Higgs doublet, $\langle H \rangle = v_{\rm SM}$, as usual, sets the scale of electro-weak symmetry breaking (EWSB).

\begin{figure}[t!!!]
\begin{center}
 \fbox{\includegraphics[height=0.18\textheight]{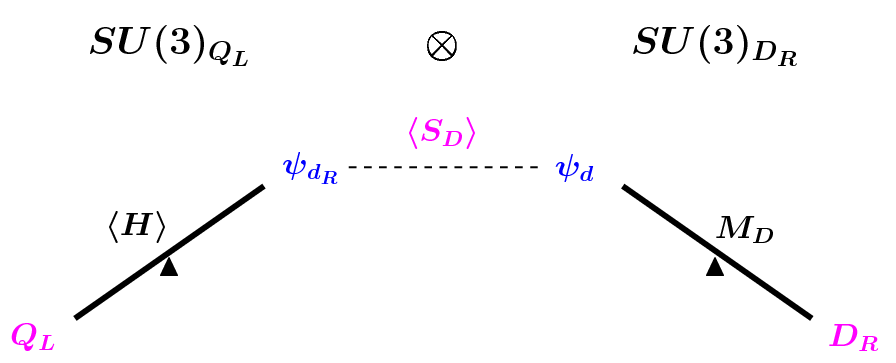}}
\end{center}
\caption{\label{illud} \footnotesize Illustration of the see-saw mechanism in the down-quark sector (see text).
The flavour spurion $S_D$ breaks the $SU(3)_{Q_L} \times SU(3)_{D_R}$ flavour symmetry
and gives masses to the heavy partners $\psi_d,\psi_{d_R}$ of the SM quarks ($Q_L$, $D_R$). 
}
\end{figure}

The different mass scales in (\ref{dim4d}) are to obey $v_{\rm SM} \ll M_{D} \ll |\langle S_{D} \rangle|$.
To derive the effective theory at the electro-weak scale, we may then simply integrate out the heavy fermions 
by solving the classical e.o.m.\ for $\psi_{d}$ (or, equivalently, $\psi_{d_R}$),
\begin{align}
 \frac{\partial {\cal L}}{\partial \psi_{d}} &= \langle S_D \rangle  \, \psi_{d_R} + M_D\, D_R \stackrel{!}{=} 0 \quad
\Rightarrow \quad \psi_{d_R} = - M_D \, \langle S_D \rangle^{-1} \, D_R \,.
\end{align}
Inserting the e.o.m.\ back into ${\cal L}$ generates effective
Yukawa terms at energies below $\langle S_{D} \rangle $,
\begin{align}
 {\cal L}_{\rm eff} &=
 - M_D \, \bar Q_L \, H \, \langle S_D \rangle^{-1} \, D_R + \mbox{h.c.} 
\end{align}
One therefore encounters an inverted relation between the down-quark Yukawa matrix and the spurion VEV $\langle S_D\rangle$,
\begin{align}
 & Y_D = M_D \,  \langle S_D \rangle^{-1}  \,.
\end{align}
This new kind of see-saw mechanism is also illustrated in Fig.~\ref{illud}:
(i) 
The left-handed SM quark doublet $Q_L$ and a new right-handed quark singlet $\psi_{d_R}$ 
transform in the fundamental representation of an $SU(3)_{Q_L}$ flavour symmetry. They couple via the SM Higgs doublet
in a gauge-invariant way under both, the SM symmetry group and the local flavour symmetry.
(ii)
The right-handed SM down-quark singlet $D_R$ and a new left-handed quark singlet $\psi_d$
transform in the fundamental representation of an $SU(3)_{D_R}$ flavour symmetry. They allow for a gauge-invariant
Dirac mass term parameterized by $M_D$.
(iii)
The two sectors of the flavour symmetry group are connected via the flavour spurion field $S_D \sim(\bar 3,3)$
which transforms in a bi-fundamental representation of  $SU(3)_{Q_L} \times SU(3)_{D_R}$ and is a singlet under the SM gauge group.
It thus couples to the new fermion fields, only. The VEV of $S_D$ breaks the $SU(3)_{Q_L} \times SU(3)_{D_R}$ 
flavour symmetry and gives masses to $\psi_{D_R}$ and $\psi_d$. 
(iv)
Upon integrating out the heavy partners of the down-type quarks, one generates an effective
Yukawa term for the SM down-quark fields with a Yukawa matrix inversely proportional to $\langle S_D\rangle$.

\subsection{Up-Quark Sector}

In \cite{Grinstein:2010ve}, it has been shown that an anomaly-free fermion spectrum for
the $SU(3)_{Q_L}\times SU(3)_{U_R} \times SU(3)_{D_R}$ flavour symmetry of the SM can
be obtained if the up-quark sector is treated analogously to the down-quark sector,
adding another set of heavy singlet fields $\psi_{u_R},\psi_u$ and a spurion field 
$S_U$ transforming as a bi-doublet under $SU(3)_{Q_L} \times SU(3)_{U_R}$.
In our model, we propose a modification, 
\begin{itemize}
 \item which also contains new heavy quark \emph{doublets}, allowing for an embedding into 
       10-plets of $SU(5)$,
 \item which is based on a \emph{different} flavour symmetry group in the quark sector, 
      $$SU(3)_{Q_L=U_R^c} \times SU(3)_{D_R}\,,$$
 \item and where, in the up-quark sector, the flavour symmetry is broken by a spurion field transforming as 
       a 6-plet of $SU(3)_{Q_L=U_R^c}$ (reflecting a symmetric up-quark Yukawa matrix in $SU(5)$). 
\end{itemize}
In this case, a see-saw mechanism can be achieved by the following terms in the dimension-4 Lagrangian,
\begin{align}
{\cal L} \ \ni \  & \phantom{+} 
\frac12 \left( \bar Q_L \, \tilde H\, \psi_{u_R}  + \bar \psi_Q \,\tilde H \, U_R \right) \cr 
& +T_U \left( \bar \psi_{u} \, \psi_{u_R} +  \bar \psi_Q \, \psi_{Q_R} \right)
+ M_U \left(\bar \psi_{u} U_R + \bar Q_L \, \psi_{Q_R} \right) + \mbox{h.c.}
\label{eq:up2}
\end{align}
Here, $\psi_Q$ and $\psi_{Q_R}$ are heavy quark doublets,
while $\psi_u$ and $\psi_{u_R}$ are heavy quark singlets.
Letting aside, for the moment, the case of the top quark, where $M_U \sim \langle T_U \rangle_{33}$ should be considered,\footnote{The case 
of the top quark in models with spontaneously broken flavour symmetry is special 
\cite{Feldmann:2008ja}. In the see-saw construction the mass eigenstates for
the top quark and its heavy partner will be linear combinations with
a mixing angle controlled by $M_U/\langle T_U \rangle_{33}$, see \cite{Grinstein:2010ve} for details.}
we may again integrate out the heavy fermions by solving the classical e.o.m.\ to obtain
the effective up-quark Yukawa matrix. Similarly as before, it is proportional to $M_U \,  \langle T_U \rangle^{-1} $,
weighted by the coupling constants that have not been shown in (\ref{eq:up2}).

\begin{figure}[t!!!]
\begin{center}
 \fbox{\includegraphics[height=0.22\textheight]{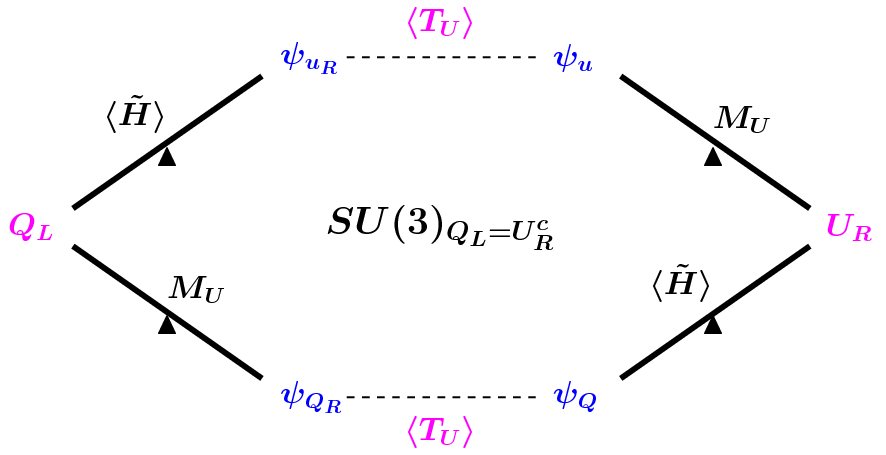}}
\end{center}
\caption{\label{illuq} \small 
Illustration of the see-saw mechanism in the up-quark sector (see text).
The flavour spurion $T_U=(T_U)^T$ breaks the $SU(3)_{Q_L=U_R^c}$ flavour symmetry
and gives masses to the heavy partners of the SM quarks. 
The fermions in the lower left (upper right) part of the figure are $SU(2)_L$ doublets (singlets).
}
\end{figure}

The situation in the up-quark sector is illustrated in Fig.~\ref{illuq}:
(i) 
The SM quark doublet $Q_L$, a new heavy right-handed doublet $\psi_{Q_R}$ and a new singlet $\psi_{u_R}$ 
transform in the fundamental representation of $SU(3)_{Q_L=U_R^c}$. The field $\psi_{u_R}$ couples to $Q_L$
via the SM Higgs doublet, while the right-handed doublet $\psi_{Q_R}$ has a Dirac mass term with $Q_L$.
(ii)
Similarly, the right-handed SM up-quark singlet $U_R$,  a new left-handed doublet $\psi_Q$ and a new singlet $\psi_{u}$
transform in the anti-fundamental representation of $SU(3)_{Q_L=U_R^c}$, with analogous couplings as in (i).
(iii)
The $SU(3)_{Q_L=U_R^c}$ flavour symmetry is broken by the VEV of the symmetric spurion matrix $T_U\sim \bar 6$
which couples the new fermion fields $\psi_{Q_R}$ and $\psi_Q$, or $\psi_{u_R}$ and $\psi_u$, respectively,
and makes them massive.
(iv)
Upon integrating out the heavy partners of the up-type quarks, one generates an effective
Yukawa term for the SM up-quark fields with a Yukawa matrix inversely proportional to $\langle T_U \rangle$.

It is instructive to derive the number of light quark fields (in the limit $\langle H\rangle \to 0$).
From the three doublet (six singlet)  Weyl spinors (for each family), two (four) Weyl spinors become massive
due to the Dirac masses generated from the scalar spurions $S_{D}, T_U$, 
leaving one light doublet and two light singlets which are identified as $Q_L,U_R,D_R$ in the SM.

\subsection{Lepton Sector}

In the lepton sector, we consider the flavour symmetry group $SU(3)_{\ell_L} \times SU(3)_{E_R} \times SU(3)_{\nu_R}$,
where the first two groups refer to the left-handed lepton doublet and right-handed singlet in the SM, and the
last factor corresponds to an additional right-handed Dirac neutrino $\nu_R$ for each family.
The masses for the SM leptons and their heavy partners descend from the dim-4 Lagrangian
\begin{align}
 {\cal L}^\ell = &
\left( \bar E_R \, H^\dagger + \bar \psi_{\nu_R} \, \tilde H^\dagger \right) \psi_\ell
 + \bar \psi_{\ell_R} \, S_E \, \psi_{\ell} + \bar\psi_{\nu_R} \, S_\nu \, \psi_\nu
  + M_E \, \bar \psi_{\ell_R}  \, \ell_L  + M_\nu \, \bar\nu_R \, \psi_{\nu} + \mbox{h.c.}
\label{option2}
\end{align}
Here, $\psi_\ell$ and $\psi_{\ell_R}$ are heavy lepton doublets which, together with the heavy quark singlets
$\psi_d$, $\psi_{d_R}$ in the down-quark sector, can be combined into 5-plets of $SU(5)$. Furthermore,
$\psi_\nu$ and $\psi_{\nu_R}$ are heavy singlet partners for the neutrinos.
We also introduced two new matrix-valued scalar spurion fields, transforming as 
$S_E \sim(3,\bar 3,1)$ and $S_\nu \sim(1,3,\bar 3)$ under the flavour-symmetry group in the lepton sector.

Solving the e.o.m.\ for $\bar\psi_{\ell_R}$ and $\psi_\nu$ in the limit $\langle S_{E,\nu}\rangle \gg M_{E,\nu}$,
one obtains
\begin{align}
 \psi_\ell &\simeq - M_E \, S_E^{-1} \, \ell_L \,,
\qquad
 \bar\psi_{\nu_R}  \simeq - M_\nu \, \bar \nu_R \, S_\nu^{-1} \,.
\end{align}
Inserting this back into$ {\cal L}^\ell$, yields the effective low-energy mass terms
\begin{align}
{\cal L}_{\rm eff} \simeq & - M_E \, \left( \bar E_R \, H^\dagger - M_\nu \, \bar \nu_R \, S_\nu^{-1} \, \tilde H^\dagger  \right) S_E^{-1} \, \ell_L \,,
\end{align}
from which we read off the Yukawa matrices responsible for the Dirac masses of charged leptons and neutrinos,
\begin{align}
 Y_E \simeq M_E \, \langle S_E\rangle^{-1} \,, \qquad 
Y_\nu \simeq - M_\nu \, \langle S_\nu\rangle^{-1} \, Y_E \simeq - M_\nu \, M_E \, \langle S_E \, S_\nu\rangle^{-1}  \,. 
\end{align}

\begin{figure}[t!!!]
\begin{center}
 \fbox{\includegraphics[height=0.18\textheight]{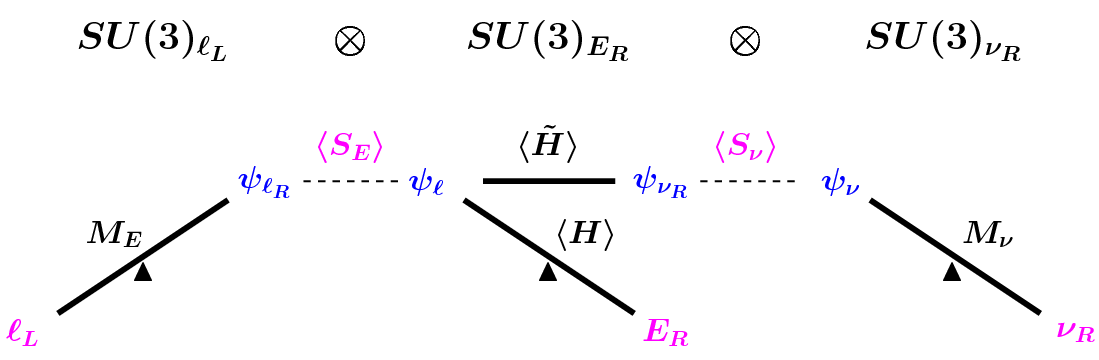}}
\end{center}
\caption{\label{illul} \small
Illustration of the see-saw mechanism in the lepton sector (see text):
The flavour spurion fields $S_E$ and $S_\nu$ break the $SU(3)_{\ell_L} \times SU(3)_{E_R} \times SU(3)_{\nu_R}$ symmetry and
give masses to the heavy partners of the SM leptons. The neutrino Dirac masses get doubly suppressed by $M_E/\langle S_E\rangle$
and $M_\nu/\langle S_\nu\rangle$. The fermions on the left-hand (right-hand) side are $SU(2)_L$ doublets (singlets).}
\end{figure}

Interestingly, the neutrino Dirac mass matrix in this case is doubly suppressed by both, $M_\nu/\langle S_\nu\rangle$ 
and $M_E/\langle S_E\rangle$, which provides a natural explanation for the smallness of the neutrino masses compared to the
masses of charged leptons. Notice, that for this purpose we did not have to introduce Majorana neutrinos.

The situation in the lepton sector is illustrated in Fig.~\ref{illul}:
(i) 
The SM lepton doublet $\ell_L$ and a new doublet $\psi_{\ell_R}$ 
transform in the fundamental representation of $SU(3)_{\ell_L}$ and couple via a fundamental
Dirac mass term, $M_E$.
(ii)
The SM singlet $E_R$, a new doublet $\psi_\ell$ and a new singlet $\psi_{\nu_R}$ transform 
in the fundamental representation of $SU(3)_{E_R}$ and couple via the SM Higgs field.
(iii)
The right-handed neutrino $\nu_R$ and its partner $\psi_\nu$ transform in the
fundamental representation of $SU(3)_{\nu_R}$ and couple via a fundamental mass term $M_\nu$.
(iv)
The flavour symmetry in the lepton sector is broken by the VEVs of the spurion matrices $S_E$
and $S_\nu$ which couple the new fermion fields from the different flavour sectors and make them
massive.
(v)
Upon integrating out the heavy partners of the leptons, one generates an effective
Yukawa term for the charged leptons with a Yukawa matrix inversely proportional to $\langle S_E\rangle$.
The neutrinos receive a Dirac mass which is doubly suppressed and stems from a Yukawa matrix
proportional to $\langle S_E S_\nu \rangle^{-1}$.

The counting of light and heavy lepton degrees of freedom works analogously as in the quark case:
From the three doublet (four singlet)  Weyl spinors for each family, two (two) Weyl spinors become massive
due to the Dirac masses generated from the scalar spurions $S_{E}, S_\nu$, 
leaving one light doublet and two light singlets which are identified as $\ell_L,E_R,\nu_R$.

\subsection{Gauged Flavour Symmetries and Anomaly Cancellation}

The transformation properties of the SM quark fields and its heavy partners 
under the $SU(2)_L$ group of the SM and under 
the $SU(3)_{Q_L=U_R^c} \times SU(3)_{D_R}$ flavour symmetry in the quark sector are as follows,
\begin{align}
 Q_L \sim 2_L(3,1)_{+y_Q} && \psi_{Q_R}^c \sim 2_L(\bar 3,1)_{-y_Q} && \psi_{Q} \sim 2_L(\bar 3,1)_{+y_Q} 
\cr 
 U_R^c \sim 1_L(3,1)_{-y_{u_R}} && \psi_u \sim 1_L(\bar 3,1)_{+y_{u_R}} && \psi_{u_R}^c \sim 1_L(\bar 3,1)_{-y_{u_R}} 
\nonumber \\[0.2em]
 D_R^c \sim 1_L(1,\bar 3)_{-y_{d_R}} && \psi_{d} \sim 1_L(1,3)_{+y_{d_R}} && \psi_{d_R}^c \sim 1_L(\bar 3,1)_{-y_{d_R}}
\label{quark_rep}
\end{align}
where we have also indicated the appropriate hyper-charges.
Concerning anomalies, it is easy to see that: 
\begin{itemize}
 \item The chiral anomalies from the new heavy quarks with respect to the SM gauge group cancel 
       between the two last columns in (\ref{quark_rep}).
 \item Concerning the mixed anomalies between the flavour symmetries and hyper-charge, we observe that
        the first two columns trivially cancel, while within the third-column the mixed anomalies cancel
        because $2y_Q = y_{u_R}+y_{d_R}$. 

\item Chiral anomalies related to the $SU(3)_{D_R}$ cancel, as we
       have one left-handed triplet and one left-handed anti-triplet.
      Chiral anomalies of the $SU(3)_{Q_L=U_R^c}$ flavour symmetry group \emph{do not} cancel by counting the
       number of left-handed triplets (2+1) and anti-triplets (2+2+1+1+1). 
       
      The anomaly-free representation in \cite{Grinstein:2010ve} could be recovered by dropping the
      heavy doublets $\psi_Q$ and $\psi_{Q_R}$.
       The alternative option that we are pursuing here\footnote{The following discussion differs from the
  one presented in an earlier preprint version of this paper and aims to resolve some of its problematic
  issues.}
  is to add two more doublet fields
\begin{align}
 \chi_L \sim 2_L(3,1)_{-5 y_Q} \,, \qquad \chi_R^c \sim 2_L(3,1)_{+5 y_Q} \,.
\end{align}
The quoted hyper-charges follow from the embedding into a 24-plet in $SU(5)$, see below.
Furthermore, in order to prevent $\chi_{L,R}$ from coupling to the other fermions, 
we introduce a discrete $Z_2$-symmetry, dubbed ``$F$-parity'' in this work,
under which $\chi_{L,R}$ are assumed to be odd, while the SM fermions and their see-saw partners are even.
The mass term for the additional fields is given by
\begin{align}
 & \bar \chi_R \, T_U \, \chi_L + \mbox{h.c.}
\label{phiLR}
\end{align}
The lightest of these states would have a mass of the order $\langle T_U\rangle_{33} \sim M_U$ and may be detectable
as a heavy quark doublet with exotic quantum numbers. 
 
\end{itemize}

In the lepton sector, a similar analysis for the $SU(3)_{\ell_L}\times SU(3)_{E_R} \times SU(3)_{\nu_R}$ symmetry 
yields 
\begin{align}
 \ell_L \sim 2_L(3,1,1)_{+y_\ell} && \psi_{\ell_R^c} \sim 2_L(\bar 3,1,1)_{-y_\ell} &&  \psi_\ell \sim 2_L(1,3,1)_{+y_\ell} 
\cr 
 E_R^c \sim 1_L(1,\bar 3,1)_{-y_{e_R}} &&  && 
\label{lepton_rep1}
\end{align}
and
\begin{align}
 \nu_R^c \sim 1_L(1,1,\bar 3)_{0} && \psi_{\nu} \sim 1_L(1,1,3)_{0} && \psi_{\nu_R}^c \sim 1_L(1,\bar 3,1)_{0}
\label{lepton_rep2}
\end{align}
where the cancellation of SM anomalies from the new leptons is obvious (one new vector representation
of charged leptons, trivial representations for right-handed neutrinos), while
the cancellation of anomalies related to each flavour sub-group can be seen by
counting the number of triplets and anti-triplets.

We have thus shown that the flavour symmetry 
$SU(3)_{Q_L=U_R^c}\times SU(3)_{D_R}\times SU(3)_{\ell_L}\times SU(3)_{E_R} \times SU(3)_{\nu_R} $
can be represented in an anomaly-free manner, and therefore we are free to introduce the corresponding
gauge bosons which will become massive by the usual Higgs mechanism as soon as the flavour spurions
get their VEVs.

\subsection{MFV Perspective}

Compared to the standard set-up of MFV \cite{D'Ambrosio:2002ex},
the up-quark sector is modified as the flavour symmetry is broken by a symmetric complex matrix $\langle T_U \rangle$ 
instead of a generic complex matrix $Y_U$. 
Counting the respective flavour parameters, we encounter 12 instead of
18 degrees of freedom in the spurion matrix, while the number of flavour symmetry generators
is reduced by 8 compared to the standard case. This leaves two additional flavour parameters 
which can be traced back to relative phases between the heavy fermionic partners of the up-quarks, 
similar as for the Majorana phases in the PMNS matrix for the neutrino sector.
In the low-energy effective Hamiltonian for $B$-, $D$-, and $K$-meson transitions one thus encounters
new sources for CP violation in addition to the SM CKM phase. 
This can be viewed as a particular example for
next-to-minimal flavour violation as defined in \cite{Feldmann:2006jk}.
Similarly, MFV in the neutrino sector \cite{Cirigliano:2005ck}, is modified as the fundamental
sources of flavour violation now transform as $\langle S_E\rangle^{-1}  \sim (3,\bar 3,1)$ and 
$\langle S_\nu\rangle^{-1} \sim (1,3,\bar 3)$, and not as the  
neutrino Yukawa matrix $Y_\nu$. Still,  the spurion fields obey consistency relations
(here between $\langle S_E S_\nu \rangle^{-1}$ and $Y_\nu$) as required in \cite{Feldmann:2006jk}. 

We emphasize that, at this point, the above considerations apply independently of a possible
embedding into a grand unified framework. From the low-energy (i.e.\ below the GUT scale)
point of view, our set-up provides an alternative realization of the idea proposed 
in \cite{Grinstein:2010ve} with a different heavy fermion spectrum, different
new sources for flavour phenomenology in the quark sector, and an interesting explanation
for small \emph{Dirac}-neutrino masses together with new sources for lepton-flavour violation.
In addition, our set-up introduces new heavy lepton and quark doublets,
which will contribute to the running of the SM coupling constant $\alpha_2(\mu)$
at energies between the scale set by the corresponding spurion VEVs
and the UV cut-off of the theory (e.g.\ $M_{\rm GUT}$).


\section{$SU(5)$ Embedding}

Our aim is to find an embedding of the heavy-fermion spectrum identified in the
previous sections in the context of a grand-unified theory based on $SU(5)$ 
fermion multiplets. The questions we want to address are:
\begin{itemize}
 \item Do the fermions fit in multiplets of $SU(5)$ ?
 \item How do the various mass terms descend from an $SU(5)$-invariant
       Lagrangian?
 \item How are the flavour symmetry groups below and above the GUT scale related?
\end{itemize}
For simplicity, we will use a symbolic
notation in terms of left-handed Weyl spinors and Higgs fields, and again suppress couplings
of ${\cal O}(1)$. The coupling terms in the Lagrangian, allowed by $SU(5)$ symmetry, 
and their decomposition into SM multiplets can, 
for instance, be found with the help of \cite{Slansky:1981yr}.
We may assume in the following a minimal Higgs sector with a $\langle 24_H \rangle$ breaking
$SU(5)$ to the SM group, and a $5_H$ containing the SM Higgs doublet responsible
for EWSB. As the flavour symmetry group above the GUT scale, we consider 
$${\cal G}_F^{\rm GUT} = SU(3)_{10}\times SU(3)_{5}\times SU(3)_1\,,$$
which is defined with respect to the GUT multiplets,  
$10_L(3,1,1)$, $\bar 5_L(1,\bar 3,1)$, $1_L(1,1,3)$,
containing the SM fermions and the right-handed neutrinos.
We further allow for scalar spurion fields 
\begin{align}
 S_5 \sim (\bar 3,3,1) \,, \qquad T_{10} \sim (\bar 6,1,1)\,, \qquad S_1 \sim(\bar 3,1,3) \,.
\end{align}

Looking at the fermion representations of the low-energy theory, as
defined in (\ref{quark_rep}) and (\ref{lepton_rep1},\ref{lepton_rep2}), we observe that
\begin{itemize}
 \item The down-quark singlets and lepton doublets, together with their
   respective heavy partners, can be combined into three 5-plets of $SU(5)$ 
   in a straight-forward manner,
\begin{align}
\mbox{SM Matter:} \qquad
&  \bar 5_L(1,\bar 3,1) \ \ni \  D_R^c,\ell_L \,,
\cr 
\mbox{Heavy Partners:} \quad
& \bar 5_L(\bar 3,1,1) \ \ni \  \psi_{d_R}^c, \psi_\ell  \,,
\cr &  5_L (1,3,1) \ \ni \  \psi_d, \psi_{\ell_R}^c \,,  
\label{5plets}
\end{align} 
where in brackets, we have given the transformation under ${\cal G}_F^{\rm GUT}$
and identified $SU(3)_{5} = SU(3)_{\ell_L^c=D_R}$ and $SU(3)_{10} = SU(3)_{Q_L=U_R^c=E_R^c}$.

\item Similarly, the right-handed neutrinos are represented by the $SU(5)$ singlets
\begin{align}
\mbox{SM Matter:} \qquad
&  1_L(1,1,3) \ = \  \nu_R^c \,, 
\cr 
\mbox{Heavy Partners:} \quad
& 1_L(3,1,1) \ = \  \psi_{\nu_R}^c \,,  
\cr  
& 1_L(1,1,\bar 3) \ = \  \psi_{\nu}  \,, 
\label{1plets}
\end{align}
with $SU(3)_1 = SU(3)_{\nu_R^c}$.

\item The up-quark sector, on the other hand, is special and -- at least in the set-up 
  identified in the last section -- prevents a straight-forward embedding into $SU(5)$.
  While the SM fermions form the usual $SU(5)$ 10-plets,
\begin{align}
\mbox{SM Matter:} \qquad
&  10_L(3,1,1)  \ \ni \  Q_L, U_R^c, E_R^c \,,  
\end{align}
  in (\ref{quark_rep},\ref{lepton_rep1}), we only encounter 
  heavy partners for the quark doublets and for the up-quark singlets, but not for
  the lepton singlets. 

  However, completing the multiplets by
\begin{align}
\mbox{Heavy Partners:} \quad
&  10_L(\bar 3,1,1) \ \ni \   \psi_Q, \psi_{u_R}^c, X_{e_R}^c  \,,  \cr 
& \overline{10}_L(\bar 3,1,1) \ \ni \  \psi_{Q_R}^c, \psi_u, X_e  \,,
\label{10plets}
\end{align}
we will see below that the additional fields $X_e$ and $X_{e_R}$ 
do not contribute to the Yukawa terms in the low-energy Lagrangian,
as long as the triplet component of the $5_H$ Higgs representation in $SU(5)$
is heavy (which requires that the usual doublet-triplet splitting problem has 
been solved in one or the other way).

\end{itemize}

With this, one can write invariant mass terms for the down-type quarks and leptons 
(in symbolic notation) as
\begin{align}
 {\cal L}^{D+E+\nu} = & 
\left( 10_L(3,1,1) \, 5_H^\dagger + 1_L(3,1,1) \, 5_H \right) \bar 5_L(\bar 3,1,1)  
\cr 
& + \left(\bar 5_L(\bar 3,1,1) \, S_5^\dagger + M_5 \, \bar 5_L(1,\bar 3,1) \right) 5_L(1,3,1) 
\cr 
& + \left(1_L(3,1,1) \, S_1 + M_1 \, 1_L(1,1,3) \right) 1_L(1,1,\bar 3) + \mbox{h.c.} 
\label{downquark}
\end{align}
which reproduces the Lagrangians in (\ref{dim4d},\ref{option2}) when decomposed into its SM components.
Integrating out the heavy fermion fields, the see-saw mechanism generates the effective
Yukawa matrices
\begin{align}
Y_D = Y_E &= M_5 \, \langle S_5\rangle^{-1} \,, \qquad 
 Y_\nu = - M_1 \, \langle S_1^\dagger\rangle^{-1} \, Y_E
\end{align}
which generates the Dirac masses for down-type quarks, charged leptons and neutrinos as indicated.

As a consequence of the unification of the charged lepton singlets with the quark doublets and
the up-quark singlets, we may write down an additional term,
\begin{align}
 {\cal L}^{\nu_R} = &  \frac12 \, 1_L(3,1,1) \, T_{10} \, 1_L(3,1,1)  + \mbox{h.c.}
\end{align}
which gives rise to a Majorana mass term for the right-handed neutrino
\begin{align}
 M_{\nu_R} = M_1^2 \, (S_1^{-1})^T \, T_{10} \, (S_1)^{-1} \,.
\end{align}
The masses for the up-quarks are realized by the following expressions,
\begin{align}
{\cal L}^{U} = & 
 \frac12 \, 10_L(3,1,1) \, 10_L(\bar 3,1,1) \, 5_H   \cr 
& \qquad + \overline{10}_L(\bar 3,1,1) \left( T_{10}^\dagger \, 10_L(\bar 3,1,1) + M_{10} \, 10_L(3,1,1) \right)
 + \mbox{h.c.} 
\label{upquark}
\end{align}
which reproduces (\ref{eq:up2}) together with the additional terms involving $X_e$ and $X_{e_R}$,
\begin{align}
 \Delta {\cal L} = & \bar X_{e_R} U_R^c \, \Phi   + \bar E_R \, X_{u_R}^c \Phi
 + \bar  X_{e_R} \, T_{10}^\dagger  \,  X_e + M_{10} \, \bar E_R \, X_e  + \mbox{h.c.}
\end{align}
When the triplet Higgs component $\Phi$ in $5_H$ is set to zero, 
the fields $X_e$ and $X_{e_R}$ can be integrated out without contributing to the effective
SM Yukawa couplings. 
Integrating out the heavy fermions from (\ref{upquark}), 
we obtain the effective SM Yukawa matrix in the up-quark sector, as before, 
\begin{align}
 Y_U = M_{10} \, \langle T_{10} \rangle^{-1} \,.
\end{align}

 Finally, we may embed the fields $\chi_{L,R}$, which we have introduced in (\ref{phiLR})
in order to compensate for the chiral anomalies in $SU(3)_{Q_L=U_R^c}$,
in a 24-plet of $SU(5)$ which is $F$-odd. This also fixes the hyper-charges quoted in  (\ref{phiLR}),
\begin{align}
 24_L(3,1,1) & \ni \chi_L(3,1,1)\,, \chi_R^c(\bar 3,1,1) \, , \ldots \qquad \mbox{($F$-odd)}
\end{align}
As indicated, this introduces a number of additional fermion fields, including a generation of SM singlets.
The 24-plet allows for a unique mass term,
$$
  24_L(3,1,1) \, T_{10} \, 24_L(3,1,1) \,,
$$
such that the lightest generation would receive a mass contribution 
of order $\langle T_{10}\rangle_{33}$. 
If -- after renormalization-group running to the weak scale -- 
the lightest of the $F$-odd states turns out to be the SM singlet, 
and $F$-parity is unbroken, it may provide a dark matter candidate.

\subsection{Anomaly cancellation}

\begin{figure}[t!!!]

\begin{center}
\begin{tabular}{c|c}
$F$-even & $F$-odd
\\
\hline  &
\\
\parbox[c]{0.50\textwidth}{\includegraphics[width=0.51\textwidth]{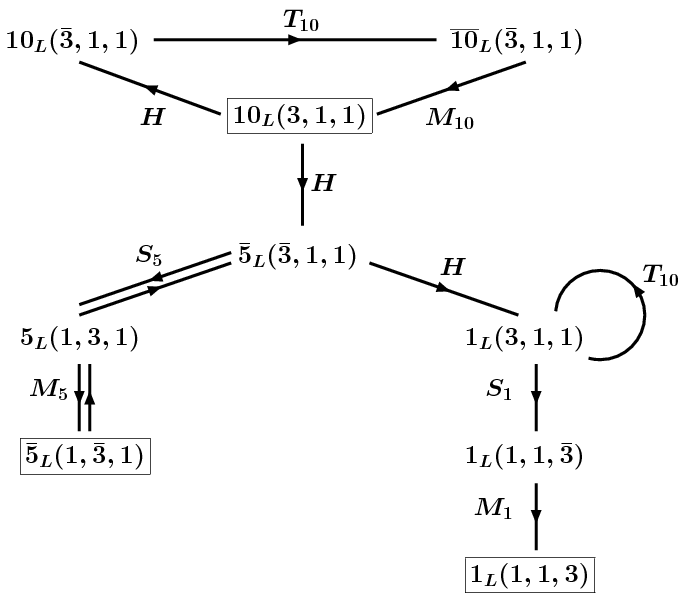}}
&
\parbox[c]{0.42\textwidth}{\includegraphics[width=0.41\textwidth]{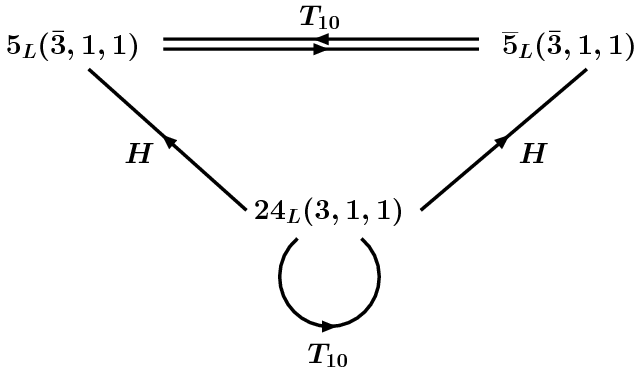}}
\end{tabular}
\end{center}
 \caption{\label{fig:summ1} \small
 $SU(5)$ fermion spectrum and realization of the see-saw mechanism. On the left-hand side we show
 the fermion representations which are even under the postulated $F$-parity. 
 The numbers in brackets indicate the transformations under 
 the GUT flavour symmetry group $SU(3)_{10} \times SU(3)_{5} \times SU(3)_1$.
 The multiplets in frames contain the light SM quarks and leptons.
 The corresponding couplings in the dim-4 Lagrangian (\ref{downquark},\ref{upquark})
 via Higgs fields $H$, fundamental mass parameters $M_{1,5,10}$, or scalar flavour spurions 
 field $S_{1,5}$, $T_{10}$ are indicated by the arrows.
 On the right-hand side we show the representations of the $F$-even fermions which are added
 to render the flavour symmetry group anomaly-free (see text).
 }

\end{figure}

Let us again discuss the anomalies associated to the fermion representations of $SU(5) \times {\cal G}_F^{\rm GUT}$.
\begin{itemize}
 \item The $SU(5)$ anomalies in the $F$-even sector cancel, because we have the same number of left-handed
       10-plets ($\overline {10}$-plets) and $\overline 5$-plets (5-plets). Notice also, that together with the
       three singlets they form complete 16-dimensional representation of $SO(10)$. The 24-plet in the $F$-odd sector
       is a real representation of $SU(5)$, and does not contribute to the anomaly.

 \item The $SU(3)_5$ flavour symmetry has a vector-like representation with one $5_L(1,3,1)$ and one $\overline 5_L(1,\bar 3,1)$, and
       the cancellation of anomalies is trivial. The same is true for the $SU(3)_1$ which contains a $1_L(1,1,3)$ and one $1_L(1,1,\bar 3)$
       fermion representation.

 \item The case with the $SU(3)_{10}$ again is more involved:
  Counting the triplet (anti-triplet) representations in the $F$-even sector, we obtain $10+1$ ($10+10+5$), leaving 
  a mismatch of $14$. The $F$-odd sector overcompensates this number by adding 24 flavour triplets. In order to obtain an anomaly-free
  spectrum with complete $SU(5)$ multiplets, the simplest option is to add $SU(3)_{10}$-triplets which are $F$-odd
\footnote{This suggests that, as an alternative to the ad-hoc F-parity, we may also consider an $SU(5)\times U(1)_X \subset SO(10)$
embedding, where the $F$-odd fermions and the Higgs fields have even $U(1)_X$ charges, and the $F$-even fermions have odd charges.} 
and  come in a vector-like representation of $SU(5)$,
$$
  5_L(\bar 3,1,1) \qquad \mbox{and} \qquad \bar 5_L(\bar 3,1,1) \qquad\qquad \mbox{($F$-odd)} \,.
$$
 Being $F$-odd, these representations can couple only among themselves via a Dirac mass term, 
$$
 5_L(\bar 3,1,1) \, T_{10}^\dagger \, \bar 5_L(\bar 3,1,1) \,,
$$
or to the 24-plet via the $5_H$ Higgs. The fermions in the $F$-odd sector thus all receive masses of
order $\langle T_U \rangle$.
\end{itemize}
The complete fermion spectrum and the various couplings are summarized in Fig.~\ref{fig:summ1}.

Our little exercise has shown that the see-saw mechanism from 
$SU(5)$ multiplets has some interesting consequences for the potential fermion spectrum 
in GUT theories which goes beyond the minimalistic approach mentioned in the introduction.
However, we shall point out a potential problem of the $SU(5)$ embedding: 
The mass parameters $M_5$ and $M_{\rm 10}$ for the 5- and 10-plets, in principle, 
can also receive contributions from the VEV of the 24-plet Higgs field which is usually 
considered responsible for $SU(5)$ breaking to the SM gauge group. This amounts to
replacing
\begin{align}
 M_{10} &\to M_{10} + \kappa_{10} \, \langle 24_H \rangle \,, 
\qquad 
 M_{5} \to M_{5} + \kappa_{5} \, \langle 24_H \rangle \,,
\label{scales}
\end{align}
and the natural scale for these parameters appears to be the GUT scale. 
The phenomenology with heavy fermion masses at or way above the GUT scale
would be quite different from (and less interesting than) 
the one discussed in \cite{Grinstein:2010ve}, 
where the masses of the lightest of the heavy fermions and gauge bosons can be close
to the TeV scale.
This calls for an alternative realization of $SU(5)$ breaking which
protects $M_5$ and $M_{10}$  against large contributions from the GUT scale
in the presence of a $24_H$ representation. 

\clearpage

\section{Summary}

We have studied local flavour symmetries in the context of $SU(5)$ unification
of the Standard Model (SM) fermion spectrum. We have shown that, below the GUT scale,
the chiral anomalies of a SM flavour symmetry group can be canceled by heavy Dirac fermions 
which receive masses from the vacuum expectation values (VEVs) of scalar spurion fields 
that break the flavour group spontaneously. 
The quark and lepton Yukawa couplings at low energies follow from 
a see-saw mechanism which -- compared to the original idea of
Grinstein, Redi and Villadoro \cite{Grinstein:2010ve} -- is extended
to the lepton sector and modified in the up-quark sector, in order
to allow for unification of the light and heavy fermions 
into $SU(5)$ multiplets.

At low energies, our construction
leads to a modification of the standard scenario considered in the context
of minimal flavour violation: Flavour transitions in the up-quark sector
are induced by a symmetric matrix  $T_U=(T_U)^T$ 
which transform as a 6-plet of the flavour symmetry $SU(3)_{Q_L=U_R^c}$.
Similarly, in the neutrino sector, the mass matrix is generated from
the \emph{product} of (small) flavour matrices, which leads to a
natural suppression of neutrino masses compared to the charged lepton masses, without the
need to introduce heavy Majorana neutrinos.
In contrast to the original model, also electro-weak doublets appear in the
heavy fermion spectrum, and therefore the unification of SM coupling constants
at the GUT scale, potentially, may be adjusted by the contributions of the new fermions
to the renormalization-group running and threshold corrections.

In order to cancel the chiral anomalies related to the flavour-symmetry group, we
have introduced a number of new fermion fields that are odd under a postulated
new $Z_2$-symmetry (dubbed ``$F$-parity in this work). From the $SU(5)$ perspective they
can be organized into a 24-plet and two 5-plets. If the lightest of these fermions
is the SM singlet contained in the 24-plet, it may serve as a dark matter candidate.
On a more speculative level, we have pointed out that our scenario
requires an alternative to the conventional mechanism to break $SU(5)$ via
a $\langle 24_H\rangle$ Higgs field, in order to protect the fundamental Dirac
mass parameters against large contributions from the GUT scale.
Answering this question goes beyond the scope of this work.

\subsection*{Acknowledgements}

I would like to thank Thomas Mannel for a critical reading of the manuscript
and for helpful discussions. I also thank Ben Grinstein for useful comments
on the issue of mixed anomalies.

\clearpage

\begin{appendix}
 
\section{See-Saw with the Pati-Salam Group}

For completeness, we also describe the embedding of the see-saw mechanism
in a GUT based on the Pati-Salam (PS) Group \cite{Pati:1973uk}, $SU(4)_C \times SU(2)_L \times SU(2)_R$.
In this case, the basic matter multiplets are
$$ 4_L\sim (4,2,1) \,, \qquad 4_R \sim(4,1,2) \,,$$
where each multiplet contains a doublet of up-, down-type quarks and leptons
with the quoted chirality. The two multiplets define the flavour symmetry group 
of the PS model,
\begin{align}
 {\cal G}_F^{\rm PS} = SU(3)_L \times SU(3)_R \,.
\end{align}
Considering again the simplest embedding of the
SM Higgs within a bi-doublet $h \sim (1,2,2)$, the see-saw Lagrangian takes
a particularly simple form,\footnote{Notice, that in (\ref{LeffPS}) we have treated the left-
and right-handed multiplet in an asymmetric way. For an explicitly left-right
symmetric version, we can simply symmetrize the effective Lagrangian, introducing
another set of heavy fermions.}
\begin{align}
 {\cal L} \ \ni \ & \bar 4_R(\bar 3,1) \, h \, 4_L(3,1) + \left( \bar 4_R(\bar 3,1) \, S_L
 + M_L \, \bar 4_R(1,\bar 3) \right) 4_L(1,3) + \mbox{h.c.}
\label{LeffPS}
\end{align}
Decomposing the above expression into its SM components, we recover the original
result for the quark sector from \cite{Grinstein:2010ve}, 
but now with two Higgs doublets and all flavour spurions equal to $S_L$
at the unification scale.
Analogous terms appear in the lepton sector, 
\begin{align}
 {\cal L} \ \ni \  & \phantom{+} \bar \ell_L \, H_2 \, \psi_{\nu_R} + \bar \psi_{\nu} \, S_\nu \, \psi_{\nu_R} + M_\nu \, \bar \psi_{\nu} \nu_R
\cr 
 & +  \bar \ell_L \, H_1 \, \psi_{e_R} + \bar \psi_{e} \, S_E \, \psi_{e_R} + M_E \, \bar \psi_{e}  E_R + \mbox{h.c.}
\label{option1}
\end{align}
with new heavy fermions singlets $\psi_\nu$, $\psi_{\nu_R}$, $\psi_e$, $\psi_{e_R}$.
In this case, the low-energy Yukawa matrices for charged leptons and neutrinos are simply
given by $Y_E \sim M_E \, \langle S_E\rangle^{-1}$ and $Y_\nu \sim M_\nu \, \langle S_\nu\rangle^{-1}$,
and we do not obtain a natural suppression of the neutrino masses as in (\ref{option2}).
Contrary to the $SU(5)$ case, we do not need additional $F$-odd fermions to get
an anomaly-free representation of the PS flavour group ${\cal G}_F^{\rm PS}$.

\end{appendix}


\begin{thebibliography}{10}



\bibitem{Pati:1973uk}
  J.~C.~Pati and A.~Salam,
  ``Unified Lepton-Hadron Symmetry And A Gauge Theory Of The Basic
   Interactions,''
  Phys.\ Rev.\  D {\bf 8}, 1240 (1973).

\bibitem{Georgi:1974sy}
  H.~Georgi and S.~L.~Glashow,
  ``Unity Of All Elementary Particle Forces,''
  Phys.\ Rev.\ Lett.\  {\bf 32} (1974) 438.

\bibitem{Carlson:1975gu}
H.~Georgi, in {\it Particles and Fields, 1974} (APS/DPF
Williamsburg), ed.\ C.~E.~Carlson (AIP, New York, 1975)
p.~575.


\bibitem{Fritzsch:1974nn}
  H.~Fritzsch and P.~Minkowski,
  ``Unified Interactions Of Leptons And Hadrons,''
  Annals Phys.\  {\bf 93} (1975) 193.

\bibitem{Mohapatra:1974gc}
  R.~N.~Mohapatra and J.~C.~Pati,
  ``A Natural Left-Right Symmetry,''
  Phys.\ Rev.\  D {\bf 11} (1975) 2558.



\bibitem{Buras:1977yy}
  A.~J.~Buras, J.~R.~Ellis, M.~K.~Gaillard and D.~V.~Nanopoulos,
  ``Aspects Of The Grand Unification Of Strong, Weak And Electromagnetic
  Interactions,''
  Nucl.\ Phys.\  B {\bf 135} (1978) 66.


\bibitem{Slansky:1981yr}
  R.~Slansky,
  ``Group Theory For Unified Model Building,''
  Phys.\ Rept.\  {\bf 79}, 1 (1981).


\bibitem{Senjanovic:2006nc}
  G.~Senjanovic,
  ``$SO(10)$: A theory of fermion masses and mixings,''
  arXiv:hep-ph/0612312.
\bibitem{Raby:2006sk}
  S.~Raby,
  ``Grand Unified Theories,''
  arXiv:hep-ph/0608183.


\bibitem{Amaldi:1991cn}
  U.~Amaldi, W.~de Boer and H.~Furstenau,
  ``Comparison of grand unified theories with electroweak and strong coupling
   constants measured at LEP,''
  Phys.\ Lett.\  B {\bf 260} (1991) 447.




\bibitem{Grinstein:2010ve}
  B.~Grinstein, M.~Redi, G.~Villadoro,
  JHEP {\bf 1011}, 067 (2010).
  [arXiv:1009.2049 [hep-ph]].


\bibitem{Albrecht:2010xh}
  M.~E.~Albrecht, Th.~Feldmann, Th.~Mannel,
  ``Goldstone Bosons in Effective Theories with Spontaneously Broken Flavour Symmetry,''
  JHEP {\bf 10} (2010) 089 [arXiv:1002.4798 [hep-ph]];
  M.~E.~Albrecht, 
  ``Two Approaches towards the Flavour Puzzle'',
   PhD Thesis, Techn.\ Univ.\ Munich (2010).



\bibitem{Feldmann:2009jung}
  Th.~Feldmann, M.~Jung and Th.~Mannel, 
  ``Sequential Flavour Symmetry Breaking,''
  Phys.\ Rev.\  D {\bf  80}, 033003 (2009)
  [arXiv:0906.1523 [hep-ph]].



%
\bibitem{D'Ambrosio:2002ex}
  G.~D'Ambrosio, G.~F.~Giudice, G.~Isidori and A.~Strumia,
  ``Minimal flavour violation: An effective field theory approach,''
  Nucl.\ Phys.\  B {\bf  645} (2002) 155
  [arXiv:hep-ph/0207036].

\bibitem{Feldmann:2008ja}
   Th.~Feldmann and Th.~Mannel,
   ``Large Top Mass and Non-Linear Representation of Flavour Symmetry,''
   Phys.\ Rev.\ Lett.\  {\bf  100}, 171601 (2008)
   [arXiv:0801.1802 [hep-ph]].
%


\bibitem{Feldmann:2006jk}
  Th.~Feldmann and T.~Mannel,
  ``Minimal Flavour Violation and Beyond,''
  JHEP {\bf 0702}, 067 (2007)
  [arXiv:hep-ph/0611095].

\bibitem{Cirigliano:2005ck}
  V.~Cirigliano, B.~Grinstein, G.~Isidori and M.~B.~Wise,
  ``Minimal flavor violation in the lepton sector,''
  Nucl.\ Phys.\  B {\bf 728} (2005) 121
  [arXiv:hep-ph/0507001].



\end{thebibliography}
\end{document}